# IOT Based Environment Monitoring System Using ESP32

Adesunmbo Adeboye Adeagbo
Dept. of Electrical and Computer Engineering
Southern Illinois University, Edwardsville
Edwardsville, USA
aadeagb@siue.edu

*Abstract*—In this project, a modern approach and technology are adopted for monitoring the environmental conditions in a particular location. This system is efficient in retrieving the environmental data from the device because the environmental conditions change spontaneously based on different atmospheric conditions. There is a need for us to pay close attention to our environment as human beings, the weather has an effect on our physical and psychological health conditions. Even if we don't consider any reason to monitor our environment, our health condition is enough to motivate us to be concerned about our environmental situation. So, environmental monitoring system makes it easier for us to have access to this data at will, by updating us about current environmental conditions and information from the weather station using apps or web pages. IoT is known as a system that connects the world together and is the way to disseminate information, so we build the system on IoT and use it as a central point for monitoring, collating, and managing our data. This project has designed an environmental system based on a microcontroller Board (EPS32) and Blynk app. This is an IoT-based project, that measures environmental data on temperature and humidity.

## I. INTRODUCTION

We are interconnected throughout the globe with the help of high-speed internet. IoT connects electronic devices and humans together, this is the way we humans, communicate with electronics that can connect through the internet. The main purpose of the Internet of Things is to connect electronic devices around the globe through the internet and retrieve the data from these devices like sensors and upload the received data to cloud service for proper analysis and informative data analysis. The devices It's equipped with environmental detectors used for dimension at any particular place and report them in real-time on the cloud.

Preface Climate change and environmental monitoring have created serious awareness that needs to be focused on from different research areas, because of their important part in mortal life, due to rapid increases in industrial factories and vehicles which have contributed negatively to the chastity of clean air and terrain. Man wants to stay streamlined about the rearmost atmospheric conditions of any place like a busy city or any other particular area.

In the earlier time, when people are home, office, or anywhere enclosed, they have little or no idea about the ecological condition outside their home or office area. If the temperature or humidity outdoors is very high or very low, how can they estimate the moistness of atmospheric conditions, they have no clue. This designed project can be adopted as an option to proffer effective solutions in this situation.

The project design followed a subsequent pattern. The first section addresses the overview of the project and relative terms, then the second section gives details about the project approach and implementation. The last section discussed the result from the analysis and conclusion respectively.

### A. Motivation

Existing weather forecasting methods were generally based on observed patterns of events, and be called pattern recognition. For example, one could observe that if the sunset was red and normal, the next day often brought nice weather. This experience gathers more than generations to produce the tradition of the time. However, not all of these predictions are reliable and since then many of them have been able to withstand rigorous statistical testing. The simplest way to predict time, persistence, depends on today's conditions to predict tomorrow's conditions.

### B. Background

Recently, a serious concern has continued to arise from the situation of the environment. Due to the increase in factories and the production of automobiles and other combustible engines, the condition of the weather continues to deteriorate, especially in agricultural areas. This condition caught the attention of many scientists and researchers to come up with a way to mitigate this deteriorating weather condition. By implementing the functionalities of the Internet of Things, many researchers have worked towards a solution for monitoring these weather conditions and have the knowledge of what is changing in the environment.

## II. RELATED WORK

The author [1] proposed an advanced monitoring system for weather based on IoT which makes it available on the internet web page. The proposed system makes use of a microcontroller (LPC2138) which is the central processing unit for all the sensors. These micro-controllers retrieve and analyze the retrieved data, which is updated to the web through the internet using GPRS. The analysis is also uploaded to the LCD display. The system consists of temperature, humidity, wind direction, and rain quantity sensors. The data processing followed 3 steps: reading the data from the sensors, storing the retrieved data through EEPROM, and finally uploading the data to the web page.

This author [2] proposed another system that is different from [1] based on the microcontroller that was used for the system. The authors measured four parameters which are temperature, humidity, moisture, and rain level sensors. All these sensors are connected to the Arduino UNO which has an in-built analog to a digital display. This Arduino UNO serves as the main processing unit, it retrieves data from all the sensors, analyzes it, and displays the output. Then finally forward the parameter to the internet through IoT to make it available everywhere.

The authors [3] proposed a system that followed a different approach by adopting NB-IoT(Narrow band Internet of Things) which enhanced the consumption power of the devices and CoAP( Constrained Application Protocol) which is designed for devices on the same constrained network. In this paper [4], the proposed design makes use of NodeMCU ESP8266-12E microcontroller with a WiFi module to retrieve data from the sensors. The microcontroller forwards the information to the server through the web, which was displayed on the web browser.

In this paper, [5] the authors designed a low-cost system for agricultural precision. ESP32 was adopted as the main controller because of its ultra-low power processing, it receives the parameter from the sensors and displays the pre-processed data on OLED. ThinkSpeak IoT was used for visualization and the visualized data were accessed by the users.

The author [6] proposed a system that monitors temperature and humidity using a DHT11 sensor chip which is used to get data for both conditions from the environment. This sensor was interfaced with Arduino UNO, which is the main processing unit. ESP822 WiFi was used to upload the data into the cloud. The system adopted IoT analytics service of ThinkSpeak and created an application for the user to monitor the conditions.

The author [7] propose a system using an ARM processor like Raspberry Pi. 5 sensors were used and a light web server framework for ARM devices was used to implement the system and a web interface was designed as well for visualization of the data.

## III. SYSTEM ARCHITECTURE

This project monitors temperature and humidity using the DHT22 sensor. This section described the approach to the project.

### A. Circuit Diagram

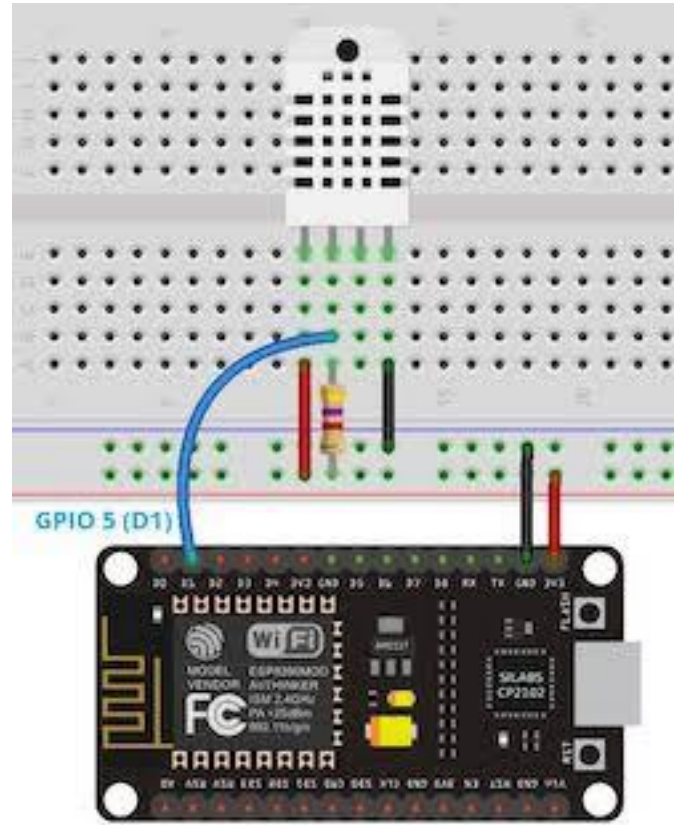

Fig. 1. Block Diagram of Environment System.

### B. Arduino

The Arduino Integrated Development Environment (IDE) is an example of an editor for writing code. It is designed using C++ language because is used mostly to write code for hardware devices. It contains libraries that have been used to communicate with Arduino hardware. The code is written in the IDE and then upload to the Arduino board. [8]

### C. Wireless Modul ESP32

ESP32 is a MCU (microcontroller unit) that is integrated Wi-Fi and Bluetooth connectivity for a wide range of applications in the control systems. It has 32-bit and has one or more

CPUs, along with memory and programmable input/output peripherals. ESP32 is capable to operate reliably in industrial environments, it can capture the temperature data, ranging from -40ºC to +125ºC. It can be used to communicate with other devices to provide bluetooth and Wi-Fi through SPI/SDIO, or I2C interfaces. [9]

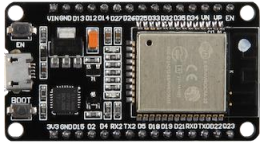

Fig. 2. ESP32.

### D. DHT22 Sensor

DHT22 is a basic which is simple to use, low-cost digital temperature and humidity sensor. The device uses a capacitive humidity sensor and a thermistor to measure the surrounding air and spits out a digital signal on the data pin. This is the advantage they have over other sensors, they are very easy to use with almost any microcontroller. This sensor can obtain new data every 2 seconds. DTH22 sensor is more precise and operates in a bigger range of temperature and humidity compared to DHT11. The sensor has four pins - VCC, DATA, NEUTRAL, and GND. [10]

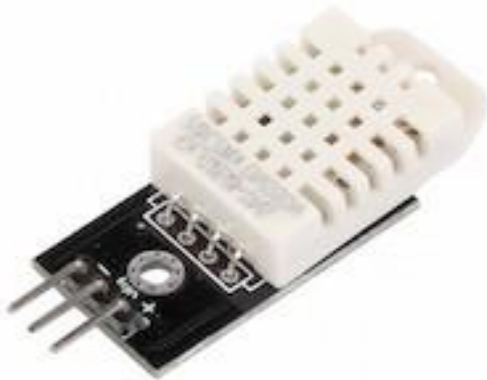

Fig. 3. DHTT22 Sensor.

### E. Connecting Cable

In connecting electrical components together, there is a need for a connecting cable. We have two categories of connecting cables, which are male and female jumper wires. Both were used in this design project.

### F. Blynk - IoT for Arduino

Blynk is a server infrastructure – the central controller of the Blynk IoT platform. Cloud is responsible for binding all the platform components together, a fully integrated suite of IoT software with many functionalities for scalability. MCUs like ESP32 and ESP8266 can be easily connected to the Blynk.Cloud to upload the data they received for analysis to give informative output. [11] [12]

### G. Assemblying the Components

We make use of Arduino Integrated Development Environment (IDE) to program the ESP32 module. The libraries for DHT22, ESP32, and Blynk were downloaded, which are been used for the functionality of the microcontroller, and DHTT22 (temperature, and humidity). To program the Blynk - IoT app interface, virtual terminals V0, and V1 are used for temperature and humidity respectively.

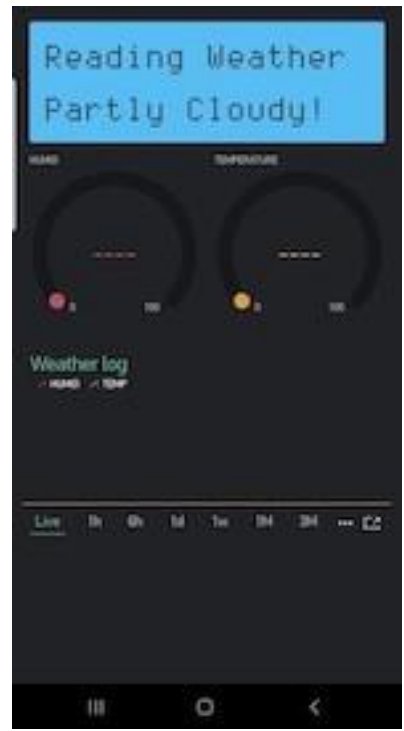

Fig. 4. Virtual terminal display settings on Blynk - IoT app

### H. Hardware Prototype

Implementing the hardware design, we make use of the ESP32 module, and temperature and humidity sensor in the design. A portable means of power was used to power up the device.

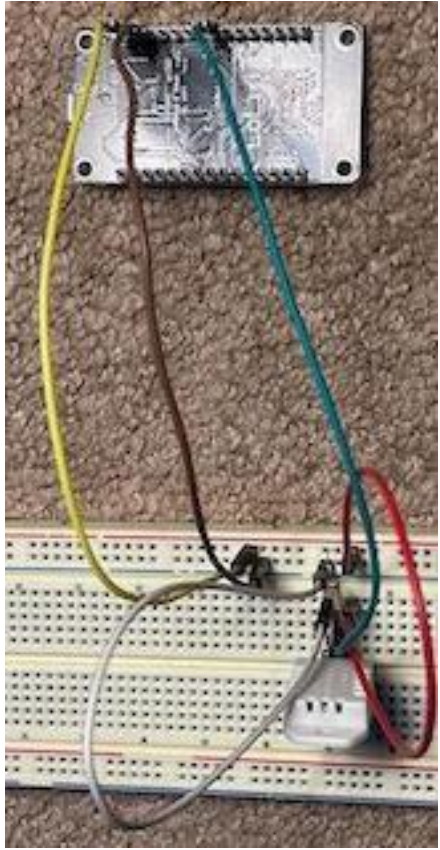

Fig. 5. Proposed model hardware implementation

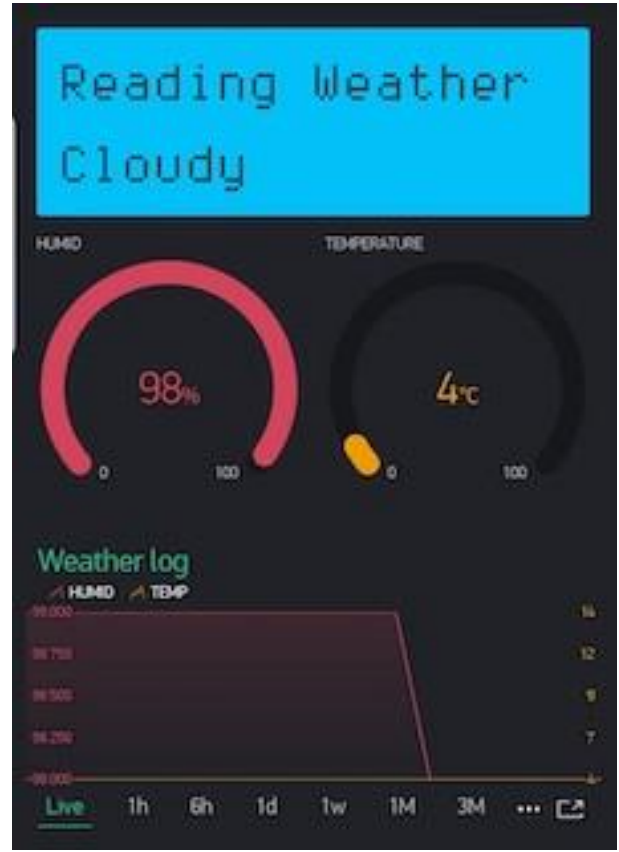

Fig. 6. The weather monitoring interface on the Blynk-IoT app

## I. Procedure

The project design follows this procedure to complete the system design.

- The first thing is for the sensors to supply the input and produce a signal on the output pin.
- The next step is for the ESP32 to retrieve output from the sensor, process it, and upload it to the cloud through a Wi-Fi connection.
- Blynk-IoT app retrieves the inputs using the virtual terminals and display

## IV. RESULT AND EVALUATION

We can see the output result from the Blynk-IoT app. The design was tested both indoors and outdoors, to get the data for the temperature and the humidity.

The output result is compared with Collinsville national weather information and the results were evaluated based on percentage error. The result is displayed in Table I.

Considering the comparison table, it can be observed that the error from the temperature is high but the humidity produces some form of good result based on accuracy and reliability. Also, I designed and programmed the Blynk app to indicate whether the weather is clear, partly cloudy, or cloudy using the information on humidity and temperature received from the sensor.

TABLE I
DATA COMPARISON OF NATIONAL INFORMATION AND EXPERIMENTAL INFORMATION ON COLLINSVILLE CITY

| Sensor | National Data | Measured Data | Percentage Error |
|---|---|---|---|
| Temperature | 6°C | 4°C | 20 |
| Humidity | 96 | 98 | 2 |

## V. CONCLUSION AND FUTURE WORK

In this project, we designed an ESP32-based Wireless environmental monitoring system which is a very low-cost design and not sophisticated. The design comes in handy with the Blynk app, which makes the weather information available to the end user at will to supply real-time weather condition information, and helps to determine how clear is the weather at that particular point in time. Considering the previous works, most of the papers actually supply the data without meaningful interpretation of what the environment actually looks like, compared to the design we proposed in this project, which actually makes use of both the temperature and humidity to determine the clarity of the weather with the data received from the sensors. The system can supply data information for temperature and humidity. The result shows some degree of accuracy and reliability except for the temperature where the percentage is a little bit higher. In the future, there is still room for improvement, especially in accommodating other atmospheric sensors like rainfall, atmospheric pressure, solar radiation, wind direction, and precipitation can be added.